\documentclass[doublecol]{epl2}
\usepackage{epsfig}
\usepackage{amsmath,amssymb,wasysym,epsfig,capt-of,ifthen,calc}
\usepackage{latexsym} %extra symbols
\usepackage{setspace} %doublespacing
\usepackage{array} %for some equations

\usepackage{delarray} %for some equations
\usepackage{afterpage} \usepackage{graphicx}% Include figure files
\usepackage{dcolumn}% Align table columns on decimal point
\usepackage{bm}% bold math \usepackage{array} \usepackage{hyperref}
\usepackage{float} \usepackage{supertabular} \usepackage{longtable}
\newcommand{\be}{\begin{equation}} \newcommand{\ee}{\end{equation}}
\newcommand{\bea}{\begin{eqnarray}} \newcommand{\eea}{\end{eqnarray}}

\usepackage{amsmath}\bibstyle{apsrev}

\title{Agglomerative Percolation in Two Dimensions}
\shorttitle{Agglomerative Percolation in Two Dimensions} %Insert here a short version of the title if it exceeds 70 characters

\author{Claire Christensen\inst{1} \and Golnoosh Bizhani\inst{1} \and Seung-Woo Son\inst{1} \and Maya Paczuski\inst{1} \\
\and  Peter Grassberger\inst{1,2} }
\shortauthor{Claire Christensen \etal}

\institute{
  \inst{1} Complexity Science Group, University of Calgary, Calgary T2N 1N4, Canada\\
  \inst{2} FZ J\"ulich, D-52425 J\"ulich, Germany, EU
}


\abstract{ We study a process termed {\it agglomerative
percolation} (AP) in two dimensions. Instead of adding sites or
bonds at random, in AP randomly chosen clusters are linked to all
their neighbors. As a result the growth process involves a
diverging length scale near a critical point. Picking target
clusters with probability proportional to their mass leads to a
runaway compact cluster. Choosing all clusters equally leads to a
continuous transition in a new universality class for the square
lattice, while the transition on the triangular lattice  has the
same critical exponents as ordinary percolation -- violating
blatantly the basic notion of universality. }

\begin{document}

\maketitle

Percolation is a pervasive concept in statistical physics and an
important branch of mathematics~\cite{kesten}. It typifies the
emergence of long range connectivity in many systems such as the
flow of liquids through  porous media~\cite{Stau-1994}, transport
in disordered media~\cite{kirkpatrick}, spread  of disease in
populations~\cite{Moor-2000},  resilience of networks to
attack~\cite{callaway},  formation of gels~\cite{Adam-1981} and
even of social groups~\cite{Mich-1966}.  It also underlies a
number of other critical phenomena --  like the Ising
order/disorder transition, which is a percolation transition on
the set of spins with given sign~\cite{Fortuin-1972}.

 The phase transition in ordinary percolation (OP), where  bonds or sites are added at random, represents a broad universality  class.  Recently   Achlioptas {\it et al.}~\cite{Achli-2009} made a simple modification by, at each step, selecting among two possibilities  the link that leads to the slowest growth of large clusters.  This global choice
introduces a large length scale -- the system size $L$ -- which
can alter universality.  Indeed, they concluded  that an unusual,
discontinuous transition (called ``explosive percolation") emerges
where   a macroscopic cluster appears suddenly while at the same
time scaling in other quantities is observed
~\cite{Ziff-2009,Ziff-2010,radicchi_explosive}.  Various
modifications of the  rule have been made~\cite{d2010local,
hans_explosive} -- all finding evidence of a discontinuous
transition.  Although the claim for discontinuity
in~\cite{Achli-2009} was refuted later  in~\cite{Costa-2009},
explosive percolation in that case does represent a new
universality class.

\begin{figure}
\begin{center}
\psfig{file=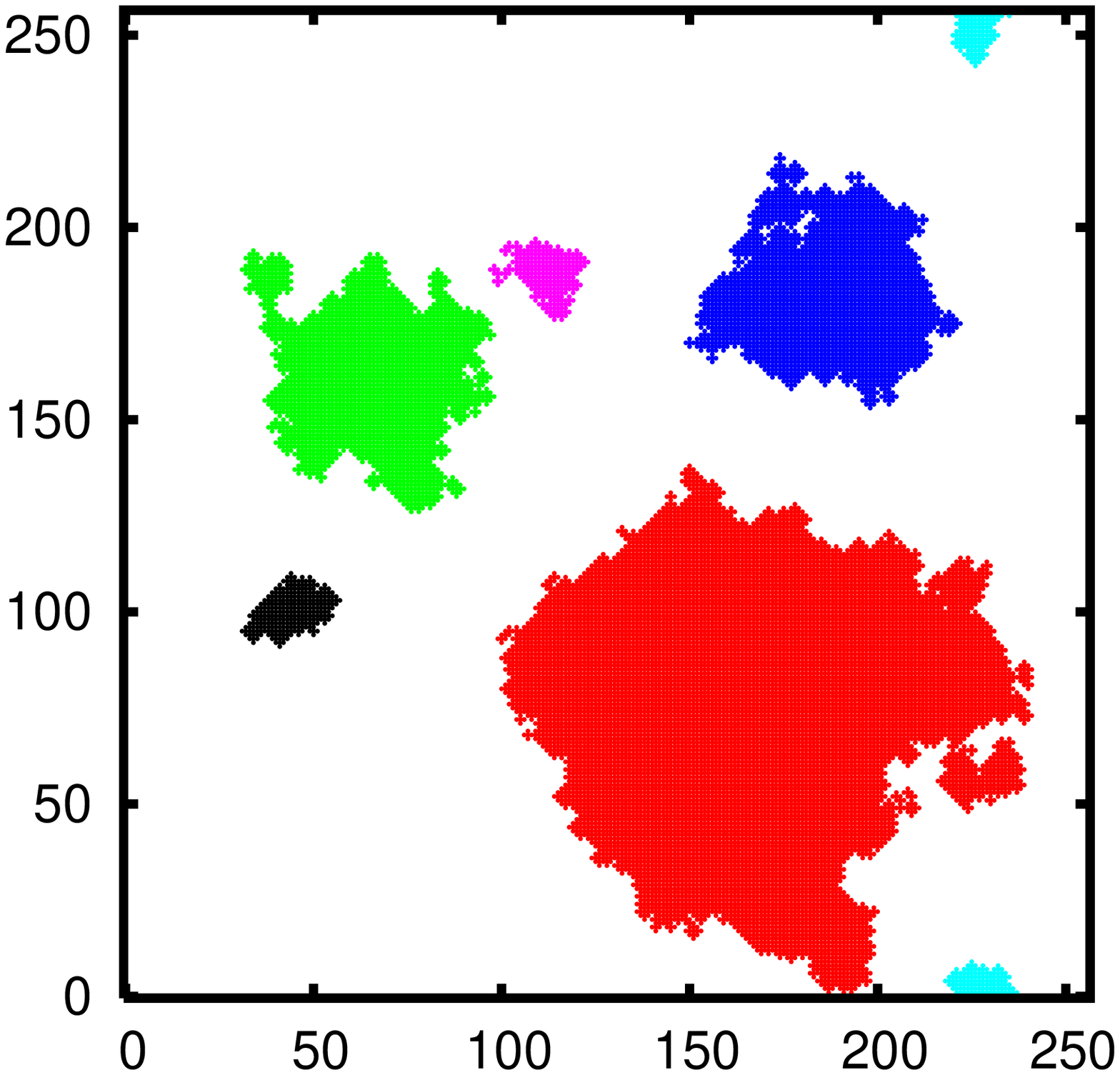,width=5.6cm, angle=270}
\psfig{file=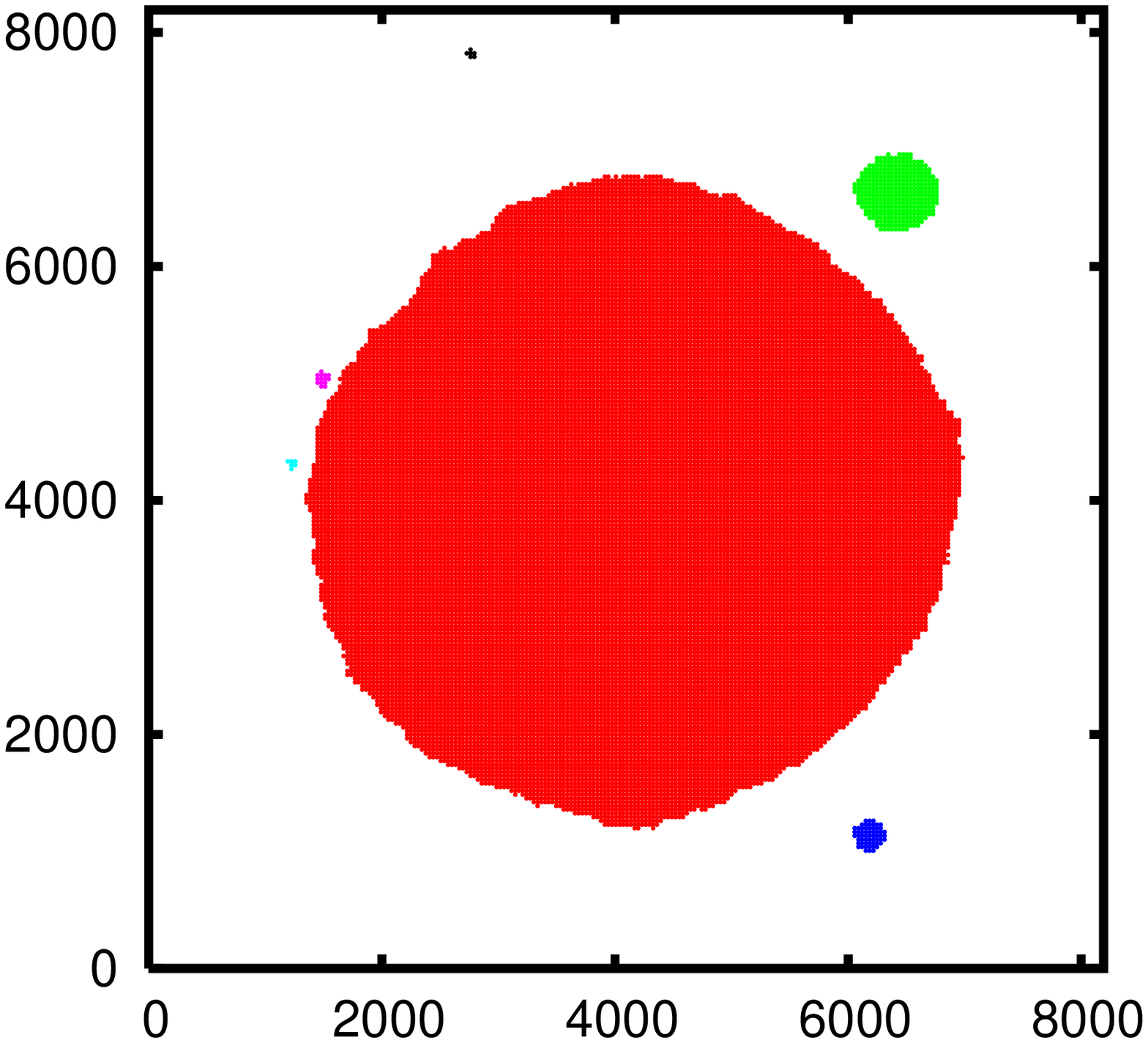,width=5.6cm, angle=270}
\caption{(Color online) The six largest clusters in mass-weighted
AP  (model (b)) on
   a square lattice with $L=256$ (top),  and $L=8192$ (bottom).
   The average cluster mass is two, or $n=N/L^2=0.5$.
   The red cluster is close to wrapping.
   All large clusters are compact.}
   \label{fig1}
\end{center}
\vskip -1 cm
\end{figure}

Here we discuss a percolation  process that also contains a
potentially large length scale in its definition, in this case the
correlation length $\xi$.   Our process has direct application to
the study of complex networks.  Instead of adding bonds  randomly,
we pick a random cluster and add bonds to its entire surface  in
order to link it to all adjacent clusters.  Starting with the
state where all clusters have size one, at each update $t\to t+1$
the process repeats until the entire lattice (or graph) is reduced
to a single cluster. We call this ``agglomerative percolation"
(AP), in  analogy with cluster growth by
aggregation~\cite{Leyv-2003}.  Thus if by chance a cluster of
length scale $\ell$ is picked, links  are added simultaneously at
distances ${\cal O}(\ell)$ apart.

AP can be analyzed on any graph.  It corresponds to random
sequential renormalization~\cite{Bizh-2010} of a network, where a
single cluster is identified as a `super'-node that is a local
coarse-graining of the graph.   In this perspective, scaling laws
seen in renormalization studies of small-world
networks~\cite{Song-2005,Goh-2006,Radic-2008,Serrano-2008,Rozen-2010}
are a consequence of an AP phase transition and {\it do not
indicate fractality of the underlying graph}~\cite{Bizh-2010,
Son-2010}.   Previously AP was studied on critical
trees~\cite{Bizh-2010} and in one dimension~\cite{Son-2010}.
Scaling laws were found both analytically and numerically --  but
no phase transition occurs since  both graphs have a topological
dimension of one.
%Analysis of  various random graphs as well as real world networks is in progress \cite{Bizh-2011}.

In order to establish the phase transition in AP,  and its
relationship to OP, we analyze it in two dimensions (which is
clearly not a fractal graph),  where many exact results for OP are
known. We consider both square and triangular lattices.   Clusters
can be chosen with equal probability, or we can make biased
choices according to the mass, radius, etc. of the clusters. Here
both (a) uniform probabilities and (b) probabilities proportional
to the cluster mass are studied. Model (b) coincides with choosing
{\it sites} uniformly, and growing the whole cluster in which they
lie.  It shows runaway behavior resembling a first order
transition with compact clusters -- as in Ref.~\cite{Janssen-2004}
(see Fig.~1).  Model (a) is more subtle. Although clusters appear
fractal (see Fig.~2) and the overall character resembles OP,
fundamental differences arise. Most conspicuous is an unexpected
difference between the two lattice types: While model (a) on the
square lattice is definitely not in the OP universality class
(e.g. the average cluster size diverges at the transition), the
triangular lattice shares the same critical exponents as OP.
We believe this violation of universality must be related to the long 
range nature of the growth process for large clusters.

We use $L\times L$ lattices, with $2^5 \leq L \leq 2^{14}$.
Boundary conditions are helical: periodic in the $y$ direction
while the right neighbor of site $(x=L,y)$ is $(x'=1,y'\equiv y
(\rm{mod}\;  L)+1)$. Diagonal bonds are added to obtain
triangular lattices.
% which produces unit cells in the shape of rectangular triangles.
%In order to obtain equilateral triangles one would have to apply an affine
%transformation that makes the overall shape of the lattice rhomboidal.
We use an algorithm based on that in~\cite{Newman-2001}, augmented
by a depth first search on the target cluster, in order to find
all its neighboring clusters. The natural control parameter in OP
is $p$, the fraction of existing bonds or sites. In AP, however, the 
number of links is not uniquely defined, although it is a version of 
(correlated) bond percolation. One might join two clusters via a single 
link, but one might also put multiple links between them. Therefore, 
in AP it is more natural to use the average cluster number per site, 
\be
   n = N/L^2=\langle m\rangle^{-1} \quad ,
\ee 
where $\langle m\rangle$ is the average cluster mass. For OP,
$n(p)$ is not analytic at $p=p_c$, but is monotonic with two
continuous derivatives. Thus one can use $n$ as the control
parameter in OP and reproduce all known scaling laws. (We checked
this explicitly; see also~\cite{Ziff-2009}).  Instead of $n$, one 
might also use the number of agglomeration events, $t$, as the control 
parameter. In agreement with ~\cite{Bizh-2010,Bizh-2011}, we found 
that $t$ is more ``noisy" than $n$ and leads to slightly less clear 
results. We analyze the distribution $P_n(m)$ of cluster masses $m$ 
in configurations with cluster density $n$, and the probability $p_{\rm wrap}(n)$,
that a cluster wraps the torus in the $y-$direction, for each $L$.

In model (b), the growth rate for a cluster of mass $m$
accelerates steeply with $m$, leading to a runaway effect. A
cluster's chance to be selected is $m/L^2$. Once chosen it grows
all  along its perimeter.  Since most of its neighbors are small,
it grows into a compact shape. For any $0< \alpha < 2$ and for
$L\to\infty$, we conjecture that the largest cluster reaches mass
$m \sim L^\alpha$ at a time when $\langle m\rangle \to 1$. This
``incipient" cluster continues to  separate in mass from the
others. It wraps  the torus when $m/L^2 = {\cal O}(1)$.  Thus an
infinite incipient cluster appears  at density $n_c\to 1$ in the
limit $L\to \infty$, while wrapping occurs much later, at
$0<n_c<1$.  Fig.~1 shows the six largest clusters in a typical run
on a square lattice for both small and large $L$ at $n=0.5$. These
snapshot were taken at  a time long past the appearance of the
incipient cluster and long before it wraps. One sees that the
giant cluster becomes more dominant over all other clusters as $L$
increases. Although convergence of $n_c\to 1$ as $L\to \infty$ is
slow, it is in perfect agreement with numerical simulation results
(data not shown). The same scenario holds for the triangular
lattice.

In OP, cluster perimeters are for large clusters proportional to
their mass. Thus, if a new bond is added at each time step, the
average growth rate of a cluster is roughly $dm/dt \propto m$. In
AP model (a) -- where clusters are picked with uniform probability
-- those chosen grow by an amount proportional to their perimeter,
so again (roughly) $dm/dt \propto m$. This leads neither to a
runaway of large clusters as in model (b) nor to the retardation
of their growth as in the Achlioptas process. Therefore, model (a)
and OP cannot be distinguished by such a crude argument and their
relationship could conceivably depend on microscopic details such
as the type of lattice.

\begin{figure}
\psfig{file=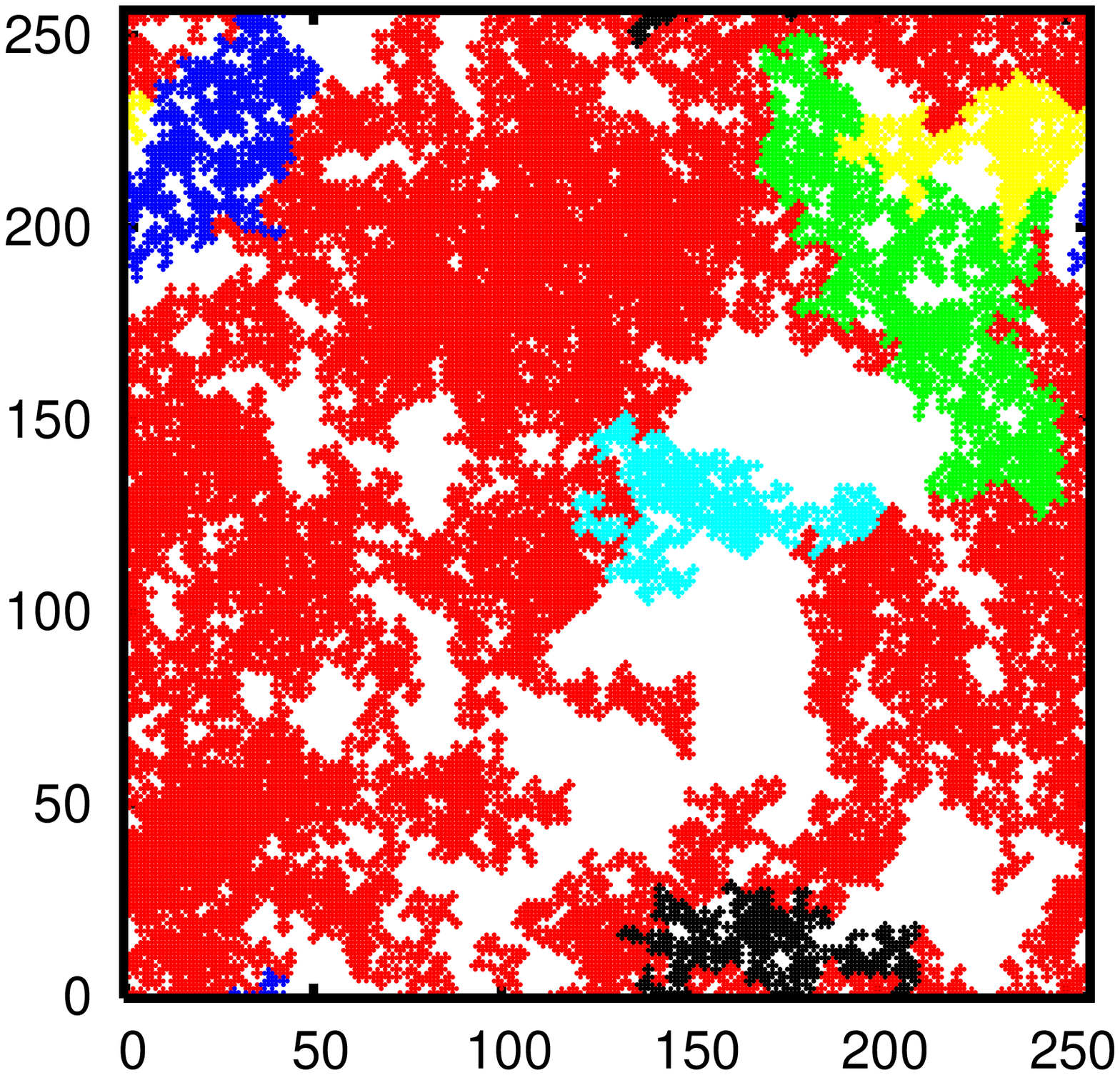,width=5.8cm, angle=270}
\psfig{file=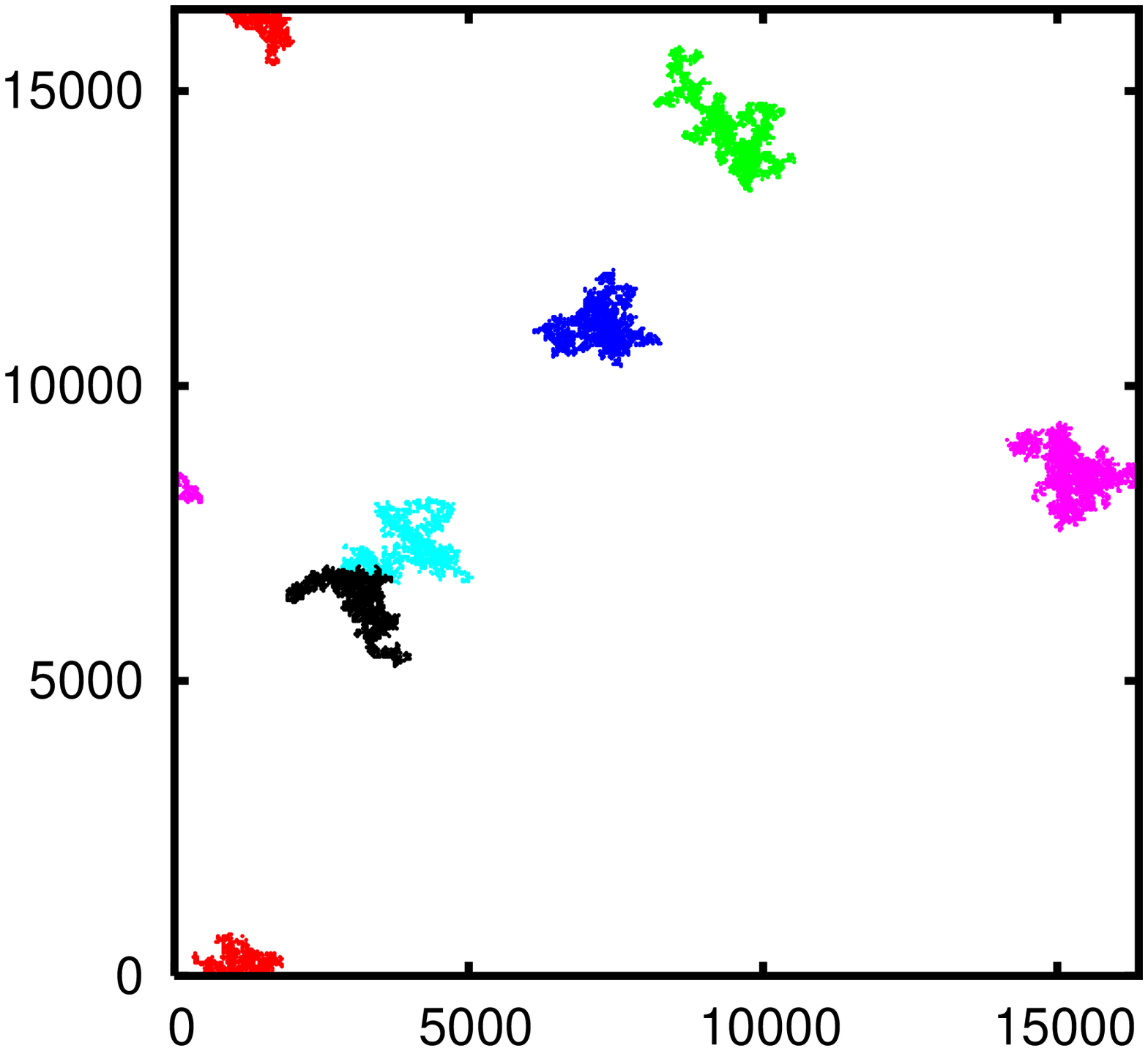,width=5.6cm, angle=270}
\caption{(Color online) The six largest clusters in typical runs
of  AP model (a) on square lattices
  for $n = N/L^2 = 0.1=1/\langle m\rangle$. Top panel: for $L=256$,
   the red cluster has already wrapped; Bottom panel: for $L=16384$, all clusters are far from wrapping. In
   both cases, clusters appear  fractal.}
   \label{fig2}
   \vskip -.5 cm
\end{figure}

For model (a), we first consider square lattices. By eye
individual configurations look like OP. However, wrapping
thresholds depend strongly on $L$. Figure~2 displays the six
largest clusters in a typical run when $n=0.1$ for both $L=256$
(top) and  for $L=16384$ (bottom). While the largest cluster
clearly wraps the small lattice, it is far from this point on the
large one. Figure~3 shows the  density $n_{c,\rm wrap}(L)$  at
which half  the runs contain a wrapping cluster. The data  fall
roughly on a straight line on a log-log plot. If deviations from a
straight line were typical finite size corrections, this would
mean that the average cluster size at the wrapping threshold
diverges as a power of $L$ when $L\to \infty$. However, for
reasons explained below, we believe that $n_{c,\rm wrap}(L) \to 0$
logarithmically as $L\to \infty$ (for explicit fits, see the
supplementary material ~\cite{supplementary-2011}). This implies that the correlation
length exhibits an essential singularity as $n\to n_c=0$.

\begin{figure}
\psfig{file=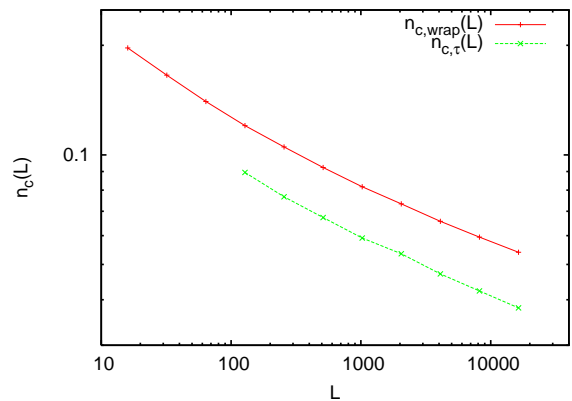,width=5.5cm, angle=270} \caption{(Color
online) Critical densities $n_c(L)$ {\it vs.} $L$ for AP model (a)
on square lattices.
    The upper curve is the average wrapping threshold. The lower one is the density at which the power
    law range in the cluster mass distribution extends furthest.  Error bars are smaller than the symbol size.}
   \label{fig3}
   \vskip -.25cm
\end{figure}

\begin{figure}
\psfig{file=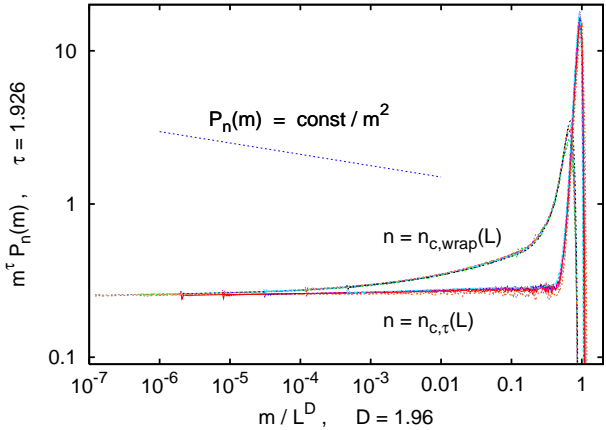,width=5.9cm, angle=270}
\caption{(Color online) Data collapse for AP model (a) on square lattices: 
   $m^\tau P_n(m)$ {\it vs.} $m/L^D$ with $\tau = 1.926$ and $D=1.96$.
   The curves with the smaller peaks are for $n=n_{c,\rm wrap}(L)$, while the others are for
   $n=n_{c,\tau}(L)$. The first are at the wrapping threshold, while the second are when
   $P_n(m)$ has the broadest power law range. The straight tilted line corresponds to
   $P_n(m)\propto m^{-2}$. System sizes are $L=256, 512, \ldots 16384$.}
   \label{fig4}
   \vskip -.5cm
\end{figure}

Mass distributions $P_n(m)$ for $n=n_{c,\rm wrap}(L)$ are
displayed in Fig.~4 using a data collapse method which compares
$m^\tau P_n(m)$ to $m/L^D$, with $\tau=1.926$ and $D=1.96$. A
perfect collapse corresponds to a finite size scaling (FSS) {\it
ansatz} 
\be
   P_n(m) = m^{-\tau} f(\psi(n,L), m/L^D) \quad ,                   \label{ansatz}
\ee 
which generalizes the standard FSS {\it
ansatz}~\cite{Stau-1994} where $\psi(n,L) = (n-n_c)L^{1/\nu}$.
Except for peak heights the data collapse is excellent. The
apparent values for $\tau$ and $D$ deviate significantly from
their values in OP ($\tau = 2.055$ and $D = 1.89$ in two
dimensions). At  $n=n_{c,\rm wrap}(L)$, $m^\tau P_n(m)$ is not
horizontal over a wide range of masses. Hence $n=n_{c,\rm
wrap}(L)$ is not equal to $n_{c,\tau}(L)$.  The latter is the
density at which a power law in $P_n(m)$ extends over the broadest
range. Curves for $m^\tau P_{n=n_{c,\tau}(L)}(m)$ are also
shown in Fig.~4 and exhibit data collapse with a remarkably wide power law range.  % for the given
%system sizes. They do not include the wrapping cluster, which for $n=n_{c,\tau}(L)$ contains most
%of the lattice sites.  Indeed
  The power law regime describes the relatively few remaining clusters
in configurations dominated by one wrapping cluster. The values
$n_{c,\tau}(L)$ are also shown in Fig.~3.  and decrease similarly
to $n_{c,\rm wrap}(L)$ as $L$ increases.

\begin{figure}[htbp]
\psfig{file=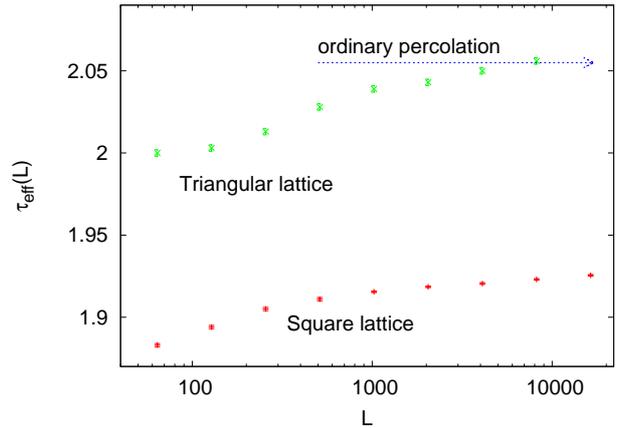, width=6.0cm, angle=270}
\caption{(Color online) Plots of $\tau_{\rm eff}(L)$, the effective Fisher exponents estimated from
   the longest stretches in $P_n(m)$ that are compatible with pure power laws. As these estimates
   are somewhat subjective, the error bars are subjective as well. But their order of magnitude is
   consistent with the smoothness of the data with varying $L$. Clearly there is non-trivial $L$
   dependence both for the triangular and for the square lattice, with $\tau_{\rm eff}(L)$ increasing
   with $L$ in both cases. While $\tau_{\rm eff}$ is compatible with the value $\tau =187/91 = 2.0549\ldots$
   in the case of the triangular lattice, it is much smaller for the square lattice.}
   \label{fig5}
\end{figure}

If $\tau \leq 2$ as suggested by Fig.~4, then the average cluster
size at criticality diverges as $L\to \infty$, in agreement with
Fig.~3 but in stark contrast to OP. Accepting this, the two
possible (scaling) scenarios are $\tau=2$ or $\tau<2$. If
$\tau<2$, $n_c(L)$ vanishes as $\sim L^{-\delta}$ with some
$\delta>0$, and apparent values for $\tau$ and $D$ should not vary
much with $L$. Neither of these statements is correct. Figure~3
shows definite curvature, and  the best fit values for $\tau$ and
$D$ both increase slightly but significantly with $L$, see Fig.~\ref{fig5}
and \cite{supplementary-2011}). While these small shifts are not 
visible on the scales shown in Fig.~4, they do not diminish as $L$ 
increases. One would not expect to see large corrections to scaling that 
could explain Fig.~\ref{fig5} if $\tau >2$ since in that case the 
average cluster size is finite.

\begin{figure}
\psfig{file=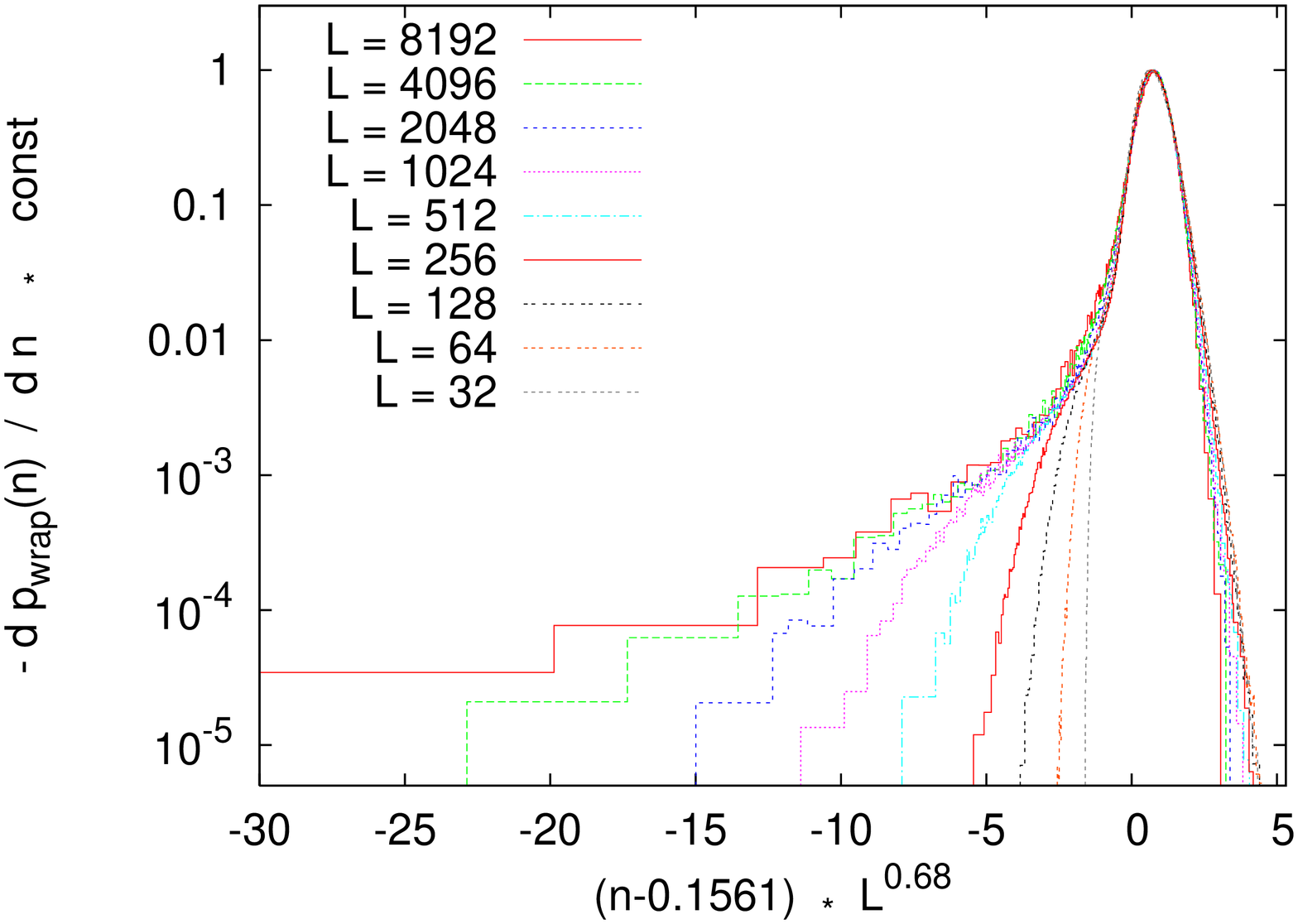,width=5.9cm, angle=270}
\caption{(Color online) Data collapse plot for the probability density 
   $dp_{\rm wrap}(n)/dn$ against $(n-n_c)L^{1/\nu}$ with $n_c=0.1561$ and 
   $\nu=1.47$ for model (a) on triangular lattices.
   The scale on the $y$-axis is adjusted such that all curves peak at $y=1$.}
   \label{fig6}
   \vskip -0.5cm
\end{figure}

Since the numerical value of $D$ is determined from the positions
of the peaks in Fig.~4, $D$ is actually the fractal dimension of
the largest cluster. The contribution to $\langle m\rangle$ from
this cluster is $s_{\rm max}/N=s_{\rm max}\langle m\rangle/L^2$. If one
assumes Eq.~(\ref{ansatz}) and $\tau < 2$, then one can show that
the largest cluster makes a non-vanishing contribution to $\langle
m\rangle$ as $L\to \infty$. This can only happen if $s_{\rm max}\sim
L^2$, showing that the largest cluster has $D=2$ if $\tau <2$.
Furthermore, one would expect convergence of the apparent $D$ to
follow  a power law in that case. But the slow convergence of $D$
from below indicates again that the behavior is dominated by
logarithms.

\begin{figure}[]
\psfig{file=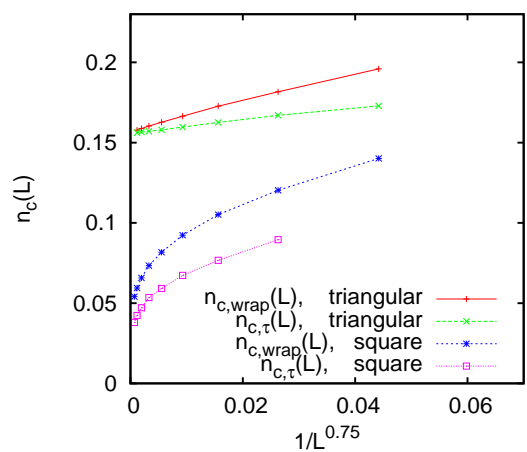,width=6.5cm, angle=270}
\caption{(Color online) Effective critical cluster densities for model (a) versus
   $L^{-3/4}$, where $L$ is the lattice size. For ordinary percolation, where $n_c(L) -n_c
   \sim L^{-1/\nu}$ with $\nu=4/3$, this should give straight lines. For each lattice type (triangular:
   upper pair of curves; square: lower pair of curves) we show results obtained with two different
   operational definitions for the critical point: (i) Maximal range of the power law
   $P_n(m) \sim m^{-\tau}$, and (ii) the probability to have a cluster that wraps around a lattice
   with helical boundary conditions is equal to 1/2. The corresponding values of $n_c(L)$ are called
   $n_{c,\tau}(L)$ and $n_{c,{\rm wrap}}(L)$. Error bars are typically of the size
   of the symbols.}
   \label{fig7}
\end{figure}

We conclude thus that the true asymptotic values are $\tau= D=2$
and $\nu=\infty$. 
In addition, we measured the exponent $\sigma$ ~\cite{Stau-1994}. 
In agreement with the scaling relation $\sigma = D/\nu$, we found 
$\sigma \approx 0$ with rather slow convergence. Thus, all scaling 
relations are (trivially) satisfied (notice that the other exponents 
give no constraint, in the present case, on the order parameter 
exponent $\beta$).
%To summarize, we can exclude the possibility  that AP model (a) on the square lattice belongs to the
%same universality class  as OP.  Our preferred scenario is $\tau=D=2$, $\sigma = 0$, and $\nu=\infty$
%with logarithmic corrections, but we cannot absolutely rule out $\tau <2$ and $ D =2$.

%This conjecture is also supported by attempts to estimate the critical exponent $\sigma$,
%defined by~\cite{Stau-1994}
%\be
 % {d\log P_n(m) \over dn} \sim m^\sigma .    \label{sigma}
%\ee
%Standard scaling theory gives $D\sigma\nu=1$, where $\nu$ is the exponent for the correlation
%length.  Logarithmic convergence in Fig.~3 would suggest that $\nu=\infty$ and thus $\sigma=0$.
%Indeed, all attempts to estimate $\sigma$ from Eq.~\ref{sigma} gave only upper bounds (typically
%$\sigma\leq 0.3$) and did not exhibit a broad scaling regime.

For the triangular lattice, clusters look like those in OP.  Both
$n_{c,\tau}(L)$ and $n_{c,\rm wrap}(L)$ converge rapidly to the
same (finite) critical value $n_c = 0.1561\pm 0.0002$. Indeed, the
best estimates of $\tau(L)$, obtained by fitting power laws to
$P_{n=n_{c,\tau}(L)}(m)$, also converge rapidly to $ \tau =
2.057\pm0.002$, in perfect agreement with OP (see Fig.~\ref{fig5}). Also $D$,
obtained from a data collapse as in Fig.~4, and the exponent 
$\sigma$ are both  within error equal to their values in OP,
although these error bars are larger than for $\tau$, see~\cite{supplementary-2011}.

Small apparent inconsistencies with OP arise when we try to
estimate $\nu$ using the scaling hypotheses $n_{c,\tau}(L) - n_c
\sim L^{-1/\nu}$ or $p_{\rm wrap}(n) = \phi[(n - n_c)L^{1/\nu}]$.
The first relation gives $\nu = 1.10\pm 0.07$, significantly
smaller than the value $\nu=4/3$ for OP. The second one gives $\nu
= 1.47\pm 0.05$ for $n \gtrsim n_c$. Taken at face value, these
estimates would exclude universality with OP. But we believe that
they are artifacts of large finite size corrections. Figure~\ref{fig6}
shows an attempted data collapse for $dp_{\rm wrap}(n)/dn$. While
the collapse is acceptable for $n>n_c$, huge tails develop for $n
\ll n_c$ as $L$ increases. For small $n$ these tails decay roughly
as $[(n_c-n)L^{0.6}]^{-\mu}$ with $\mu \approx 1.5$.
% To understand them, we have to define
%$dp_{\rm wrap}(n)/dn$ precisely: It is the probability that wrapping did occur for $N=nL^2$,
%but not for any larger $N$.
We checked explicitly that the tails result from events where
wrapping happened when a large target cluster was hit. In such
cases, $N$ can make huge jumps, so that the largest $n$ at which
the cluster wraps is far below the actual threshold. Thus we
conclude that AP on the triangular lattice is in the OP
universality class, with the caveat that the definition of $n$ is
affected by occasional large jumps which do not modify the main
critical exponents but which do modify the tails of scaling
functions  as in Fig.~\ref{fig6}. 

For an appreciation how different the 
behaviors are on the square and triangular lattices, we show in 
Fig.~\ref{fig7} effective critical cluster densities. We plot them 
against $L^{-3/4}$, since this should give for OP straight lines, according
to the FSS ansatz. While the data for the triangular lattice indeed follow
roughly straight lines and give a finite non-zero value of $n_c$, the same is 
definitely not true for the square lattice: Those curves strongly bend down for 
$L\to\infty$, suggesting that $n_c=1/\nu=0$. Whether the latter is correct or 
not, this figure should leave no doubt that the square lattice model is not 
in the ordinary percolation universality class.

In summary, we have studied agglomerative percolation (AP) in two
dimensions. This class of models is equivalent to random
sequential renormalization schemes~\cite{Bizh-2010} first
introduced to scrutinize the supposed
fractality~\cite{Song-2005,Goh-2006,Radic-2008,Serrano-2008,Rozen-2010}
of real world -- in particular, small-world -- networks. Regular
lattices were chosen for two reasons: (1) They are not fractal;
(2) Detailed comparison can be made with exact results for
ordinary percolation (OP). Our results display some of the rich
behavior possible in this general class of models and indicate
that at least some of the scaling behavior found
in~\cite{Song-2005,Goh-2006,Radic-2008,Serrano-2008,Rozen-2010} is
due to AP rather than any supposed fractality of the underlying
graph. If clusters are chosen with a bias for larger mass (model
(b)), a runaway effect separates the largest cluster from the
others and the behavior is completely different from OP. If
clusters are chosen with equal probability (model (a)), then only
a detailed numerical scaling analysis shows that AP is not in the
OP universality class on the square lattice. On the other hand, AP
on the triangular lattice shares critical exponents with OP.

AP may have applications beyond network renormalization. Growing
clusters appear in many different physical situations. It could
happen that further growth is triggered by some excitation where
the entire cluster suddenly invades neighboring clusters at its
boundary.  Agglomerative percolation could also describe the
growth of countries or urban areas. Countries often grow by
overrunning and incorporating neighbors during aggressive periods,
when they attack and incorporate simultaneously several of their
neighbors.
%In that case AP would mean ``aggressive percolation".

Acknowledgement: We thank Bob Ziff for  correspondence that helped
us  sharpen some of our arguments.

%%%%%%%%%%%%%%%%%%%%%%%%%%%%%%%%%%%%%%%%%%%%
\bibliographystyle{eplbib}
\bibliography{EPL_bibliography}
%%%%%%%%%%%%%%%%%%%%%%%%%%%%%%%%%%%%%%%%%%%%
\end{document}

\title{Supplementary Material for ``Agglomerative Percolation in Two Dimensions"}
\author{Claire Christensen} \affiliation{Complexity Science Group, University of Calgary, Calgary T2N 1N4, Canada}
\author{Golnoosh Bizhani} \affiliation{Complexity Science Group, University of Calgary, Calgary T2N 1N4, Canada}
\author{Seung-Woo Son} \affiliation{Complexity Science Group, University of Calgary, Calgary T2N 1N4, Canada}

\author{Maya Paczuski} \affiliation{Complexity Science Group, University of Calgary, Calgary T2N 1N4, Canada}
\author{Peter Grassberger} \affiliation{Complexity Science Group, University of Calgary, Calgary T2N 1N4, Canada} \affiliation{IAS, JSC, FZ J\"ulich, D-52425 J\"ulich, Germany}
\date{\today}

\maketitle

In this supplementary material, we present several plots. They illustrate claims for which in the 
main paper either the actual data were not shown or they were plotted in different ways.   
The present plots are to convince the reader that: 
\begin{itemize}
\item Model (b) (where clusters are selected for agglomeration according to their mass)
    develops runaway clusters very early in the process, leading 
    to a ``phase transition" at $\langle m \rangle = 1$ in the infinite system limit $N\to\infty$ (here, 
    $\langle m \rangle$ is the average mass per cluster) (Fig. S1); 
\item The same is not true for model (a), but model (a) shows qualitatively different behaviors 
    on the square and triangular lattices (Figs. S2 to S5);
\item In particular, while model (a) on the triangular lattice is in the ordinary percolation 
    universality class, it is definitely not in that universality class
    on the square lattice;
\item For the square lattice, {\it superficial analyses} would give a Fisher exponent $\tau$ and a fractal dimension
    $D$ that are both smaller than 2, but a more careful analysis, combined with analytic arguments,
    leads to the conclusion that $\tau=D=2$ exactly.
\end{itemize}

\begin{figure}[htbp]
\psfig{file=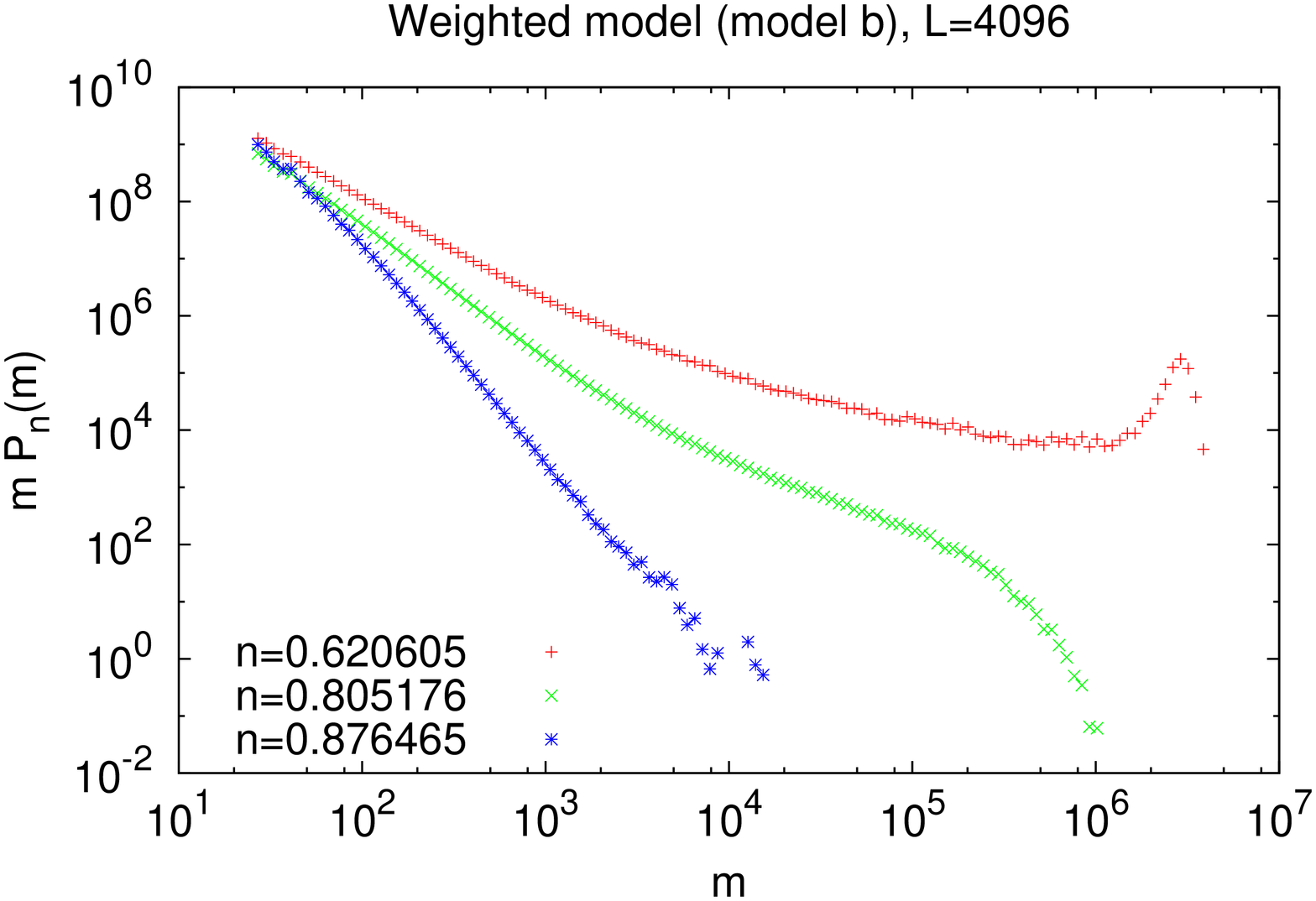,width=9.9cm, angle=270}
\caption{(Color online) Mass distributions for model (b) with $L=4096$. Each curve corresponds to 
   a fixed value of $n$ (number of clusters per site), corresponding to an average cluster mass
   $\langle m \rangle = 1/n$. The uppermost (red) curve shows a clear narrow peak, corresponding to a single
   cluster that involves a substantial (and thus not strongly fluctuating) fraction of all sites.
   The middle (green) curve shows a broad bump at its upper end, corresponding still to a runaway cluster
   in most realizations, albeit with a much wider mass range centered typically at a few percent of 
   the total mass. The lowest (blue) curve corresponds to $\langle m \rangle \approx 1.14$, i.e. by far most 
   clusters consist of a single site. In spite of this, it shows a long tail implying that {\it some}
   clusters have reached size $10^4$ and beyond. In addition to Fig.1 in the main paper, this shows
   most directly the existence of runaway clusters at values of $\langle m \rangle$ that converge to
   1 as $N\to\infty$. }
   \label{fig1}
\end{figure}

\begin{figure}[htbp]
\psfig{file=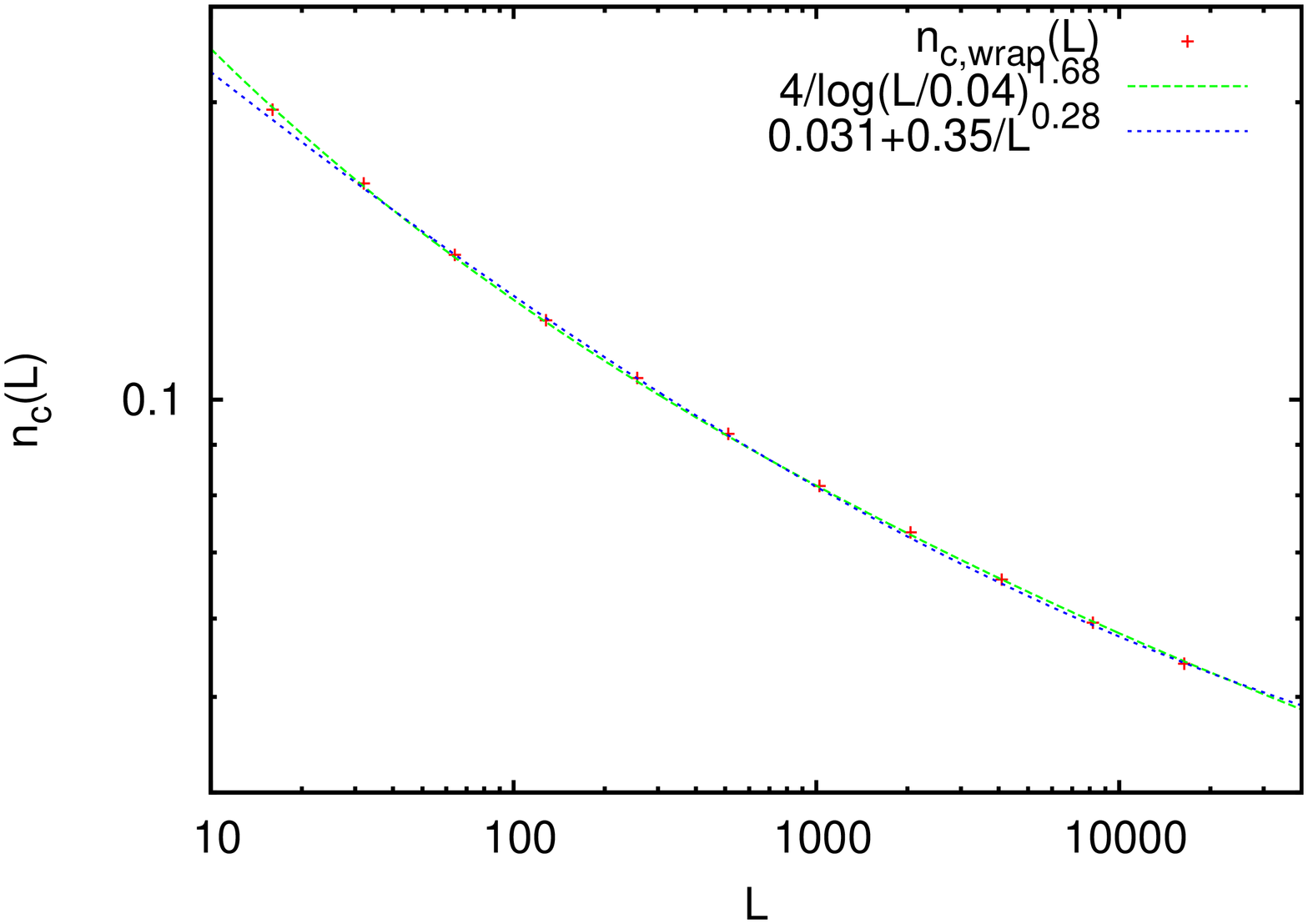,width=9.9cm, angle=270}
\caption{(Color online) Plot of $n_{c,{\rm wrap}}(L)$ versus $\log L$ for model (a) on the square lattice, 
   together with two typical fits.
   One fit (green curve) is logarithmic, $n_{c,{\rm wrap}}(L) = 4/\ln(25 L)^{1.68}$. It assumes that
   $n_{c,{\rm wrap}}\equiv n_{c,{\rm wrap}}(L=\infty) = 0$. The other fit (blue) assumes that $n_{c,{\rm wrap}}$
   is non-zero, and that $n_{c,{\rm wrap}}(L)$ has power law finite size corrections: $n_{c,{\rm wrap}}(L) = 
   0.031+0.35/L^{0.28}$. Notice that both fits involve three free parameters. The logarithmic fit is slightly 
   better, but it is not clear how significant this is, since one should expect an infinite series 
   of further correction terms in any such fit. Notice also that $n_{c,{\rm wrap}}$ is very small (i.e., 
   the average cluster mass at criticality is huge) in the second fit, in striking contrast to other (bond, site)
   percolation models on any 2-d lattice. In our opinion, this is a stronger argument against the second
   fit than the quality of the fit itself. Finally, no decent fit is possible with a pure power law 
   $n_{c,{\rm wrap}}(L) = a/L^\alpha$, expected if $\tau < 2$ -- although fits with two or more powers
   with different exponents would be possible. }
   \label{fig2}
\end{figure}

\begin{figure}[htbp]
\psfig{file=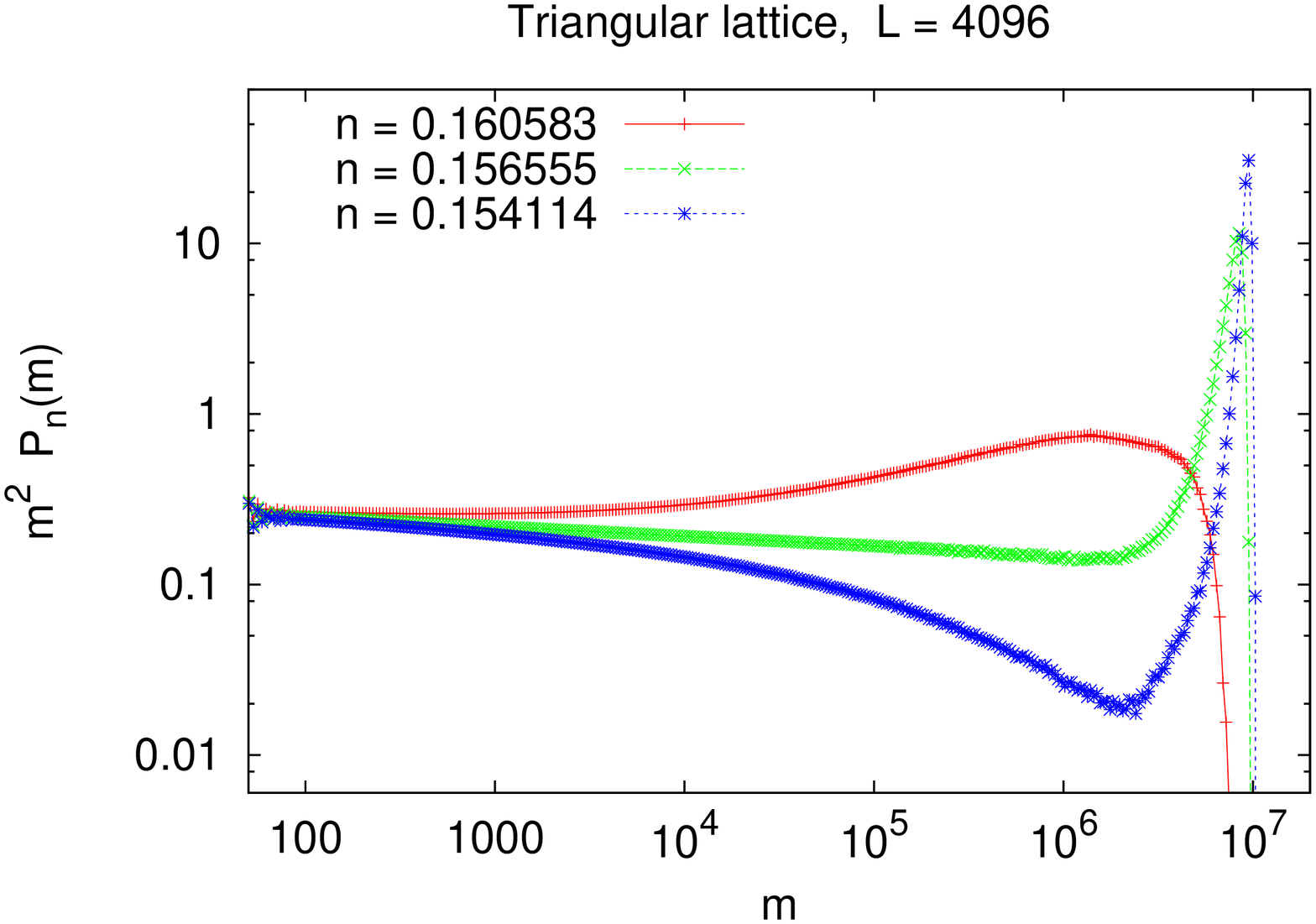,width=9.7cm, angle=270}
\caption{(Color online) Mass distributions (log-log) for model (a) on the triangular lattice with $L=4096$, 
   for three different values of $n$. The middle (green) curve has the longest straight piece and is thus 
   used to define $n_{c,\tau}$. The other two curves show the sub- (super-)critical behavior. For
   increased significance, we actually do not show $P_n(m)$ but rather $m^2P_n(m)$. The fact that 
   the central (green) curve has a negative slope clearly indicates $\tau>2$. Notice that this conclusion
   is rather robust and would not require a very precise estimate of $n_{c,\tau}$.}
   \label{fig3}
\end{figure}

\begin{figure}[htbp]
\psfig{file=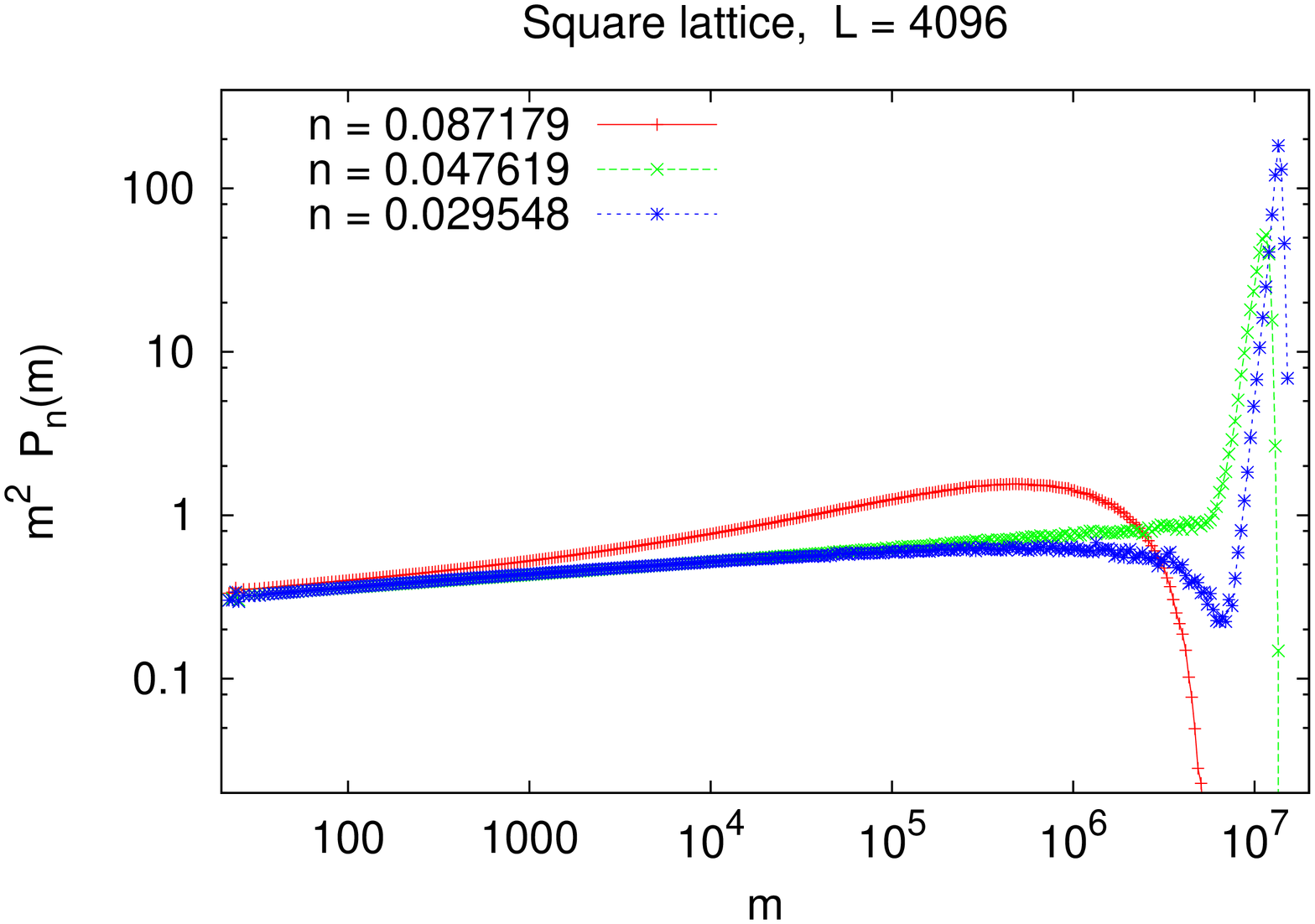,width=9.7cm, angle=270}
\caption{(Color online) Same as Fig. S3, but for the square lattice. Superficially, the graph looks 
   similar to Fig. S3, but the slope of the critical (green) curve now is positive, clearly suggesting
   that $\tau<2$ unless there are very strong corrections to scaling.}
   \label{fig4}
\end{figure}

\begin{figure}[htbp]
\psfig{file=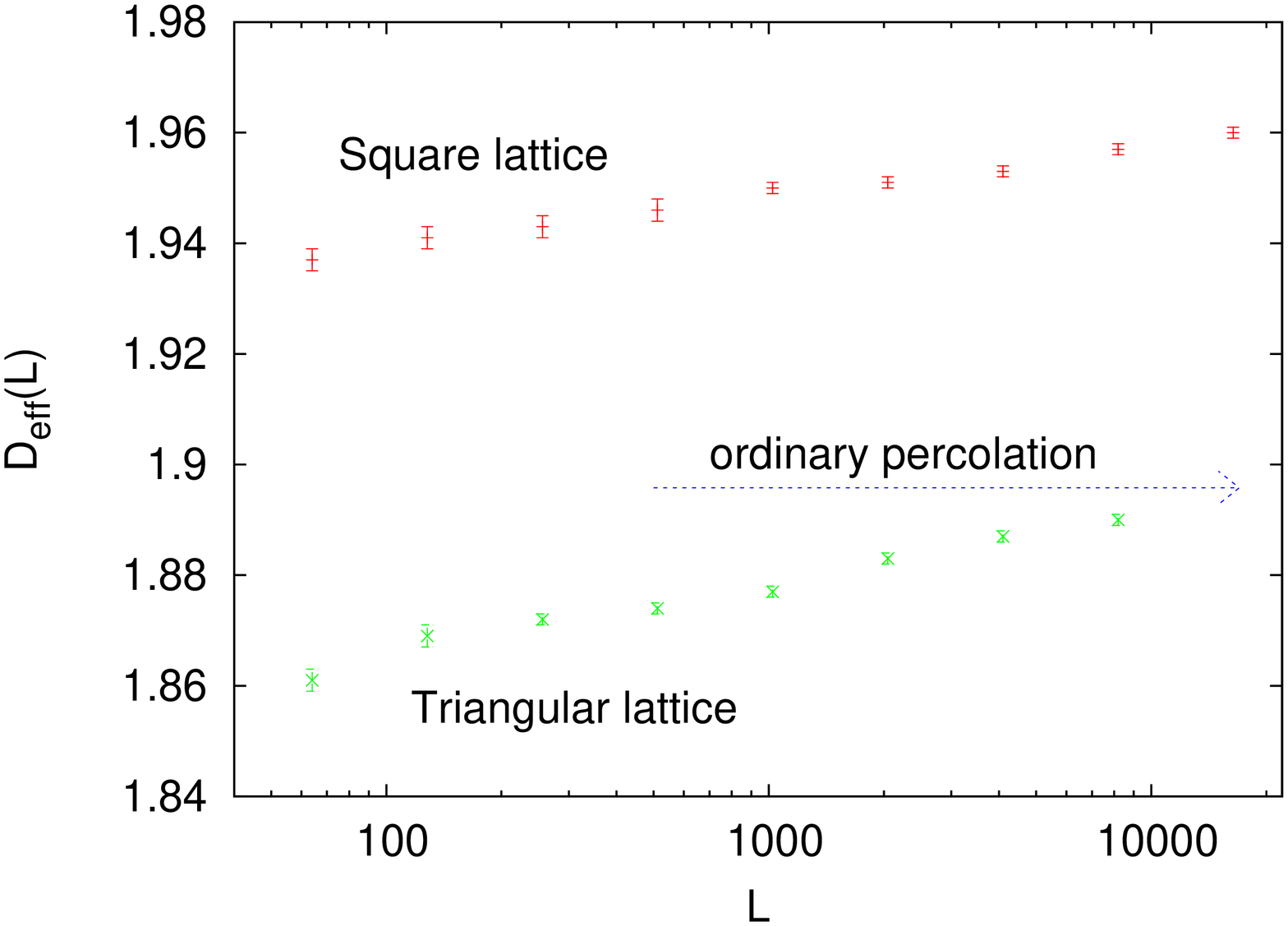,width=8.3cm, angle=270}
\caption{(Color online) Similar to Fig. 5 of the main paper, but for the fractal dimension of the largest (percolating)
   cluster. It is obtained from the positions of the peaks of the central (critical) curves in mass 
   distribution plots such as those in Figs. S4 and S5. Again we see that both curves increase with $L$. Again
   the triangular lattice data are compatible with ordinary percolation, $D = 91/48=1.8958\ldots$, but
   the square lattice data are not. There, the percolating cluster has a distinctly larger dimension 
   which is moving {\it away} from the ordinary percolation value. 
   The analytical arguments given in the main paper allow only three possible scenarios: (i) $D <2, 
   \tau>2, \nu<\infty$, and power law corrections to scaling; (ii) $D=\tau=2, \nu=\infty$, and 
   logarithmic corrections; and (iii) $D=2, \tau<2, \nu<\infty$, and power law corrections to scaling. 
   Scenario (i) (i.e., $\tau>2$) seems definitely ruled out for the square lattice by the data. 
   Scenarios (ii) and (iii) are both possible, but would require very large and slowly vanishing 
   finite size corrections for $D_{\rm eff}(L)$. At face value, the data for $n_c(L)$ shown in 
   Fig. S2 and in Fig. 3 of the main paper would favor scenario (ii), while $\tau_{\rm eff}(L)$ (Fig. 5)
   would favor scenario (iii). Our final preference of scenario (ii) is based on the fact that it 
   is the only one which can naturally accommodate logarithmic (and thus very slowly vanishing) finite 
   size corrections.}
   \label{fig5}
\end{figure}